\documentclass[10pt,preprintnumbers,showpacs,amsmath,amssymb,floatfix,twocolumn,prl]{revtex4}   
\usepackage{graphicx}
\usepackage{dcolumn}
\usepackage{bm}
\usepackage{longtable}
\usepackage{graphics}
\usepackage{amssymb}
\usepackage{amsmath}
\usepackage{xspace}
\usepackage{epsfig}
\newcommand {\NaH}{Na$_x$CoO$_2\cdot y$H$_2$O }

\newcommand {\CCC}{Cs$_2$CuCl$_4$ }

\newcommand {\dmit}{$\beta'$-[Pd(dmit)$_2$]$_2X$ }
\newcommand {\dmitn}{$\beta'$-[Pd(dmit)$_2$]$_2X$}

\newcommand {\Br}{$\kappa$-(ET)$_2$\-Cu\-[N(CN)$_2$]\-Br }
\newcommand {\Brn}{$\kappa$-(ET)$_2$\-Cu\-[N(CN)$_2$]\-Br}
\newcommand {\Cl}{$\kappa$-(ET)$_2$\-Cu\-[N(CN)$_2$]\-Cl }

\newcommand {\bI}{$\beta$-(ET)$_2$\-I$_3$ }

\newcommand {\CN}{$\kappa$-(ET)$_2$\-Cu$_2$\-(CN)$_3$ }
\newcommand {\CNn}{$\kappa$-(ET)$_2$\-Cu$_2$\-(CN)$_3$}

\newcommand {\ibid}{{\it ibid}. }
\newcommand {\etal}{{\it et al}. }

\newcommand {\etalc}{{\it et al}., }
\begin{document}
\title{Symmetry of the superconducting order parameter in frustrated systems determined by the spatial anisotropy of spin correlations}
\author{B. J. Powell}
\affiliation{Department of Physics,
University of Queensland, Brisbane, Queensland 4072,
Australia}\author{Ross H. McKenzie}
\affiliation{Department of
Physics, University of Queensland, Brisbane, Queensland 4072,
Australia}

\pacs{}

\begin{abstract}
We study the resonating valence bond (RVB) theory of the
Hubbard-Heisenberg model on the half-filled anisotropic triangular
lattice (ATL). Varying the frustration changes the wavevector of
maximum spin correlation in the  Mott insulating phase.
 This, in turn, changes the symmetry of the
superconducting state, that occurs at the boundary of the
Mott insulating phase. We propose that this physics is realised in
several families of quasi-two-dimensional organic superconductors.
\end{abstract}

\maketitle

One of the major themes in condensed matter physics over the last
few decades has been the deep connection between magnetism and
unconventional superconductivity. This is one of the key ideas that
has emerged from the study of the cuprates \cite{LeeRMP}, ruthenates
\cite{SRO}, cobaltates \cite{NCO,Ogata}, heavy fermions
\cite{Mathur}, organic superconductors \cite{springer-review},
 $^3$He \cite{Leggett}, and ferromagnetic superconductors
\cite{fmsc}. In the cuprates $d_{x^2-y^2}$ symmetry
superconductivity emerges from the doping of a Mott insulator with
N\'eel order. Many theories
\cite{AndersonRVB,LeeRMP,Pines-Scalapino,Norman}, including RVB,
suggest that in the metallic state spin correlations which are
maximal near the wavevector $(\pi,\pi)$ mediate superconductivity.
In RVB theory \cite{AndersonRVB,LeeRMP} superconductivity arises
from the same strong correlations that give rise to
antiferromagnetism in the Mott insulator. Alternative theories of
the cuprates emphasize instead the role of different physics, such
as stripes, phase fluctuations, or orbital currents \cite{Norman}.

\begin{figure}
\begin{center}
\epsfig{file=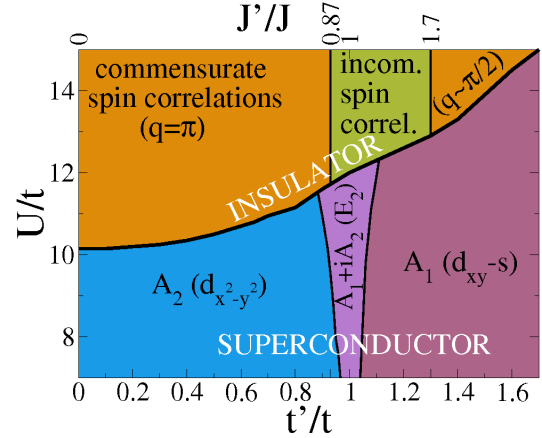,width=7cm,angle=0}
\end{center}
\caption{(Color online.) Phase diagram of the Hubbard-Heisenberg
model on the half filled ATL. As the frustration, $J'/J=(t'/t)^2$,
is varied the spin correlations change from commensurate $(\pi,\pi)$
characteristic of the square lattice for $t'/t<0.93$, to
incommensurate $(q,q)$ in the highly frustrated regime, to
commensurate $(\pi/2,\pi/2)$ for 
$t'/t\gtrsim1.3$ 
characteristic of weakly coupled chains (see Fig.
\ref{fig:ordering}). The spin correlations mediate
superconductivity, and the changes in the spin correlations cause
changes in the symmetry of the superconducting state, which changes
from `$d_{x^2-y^2}$' ($A_2$) for small $t'/t$ to `$d+id$'
($A_1+iA_2$) for $t'\sim t$ to `$s+d_{xy}$' ($A_1$) for large
$t'/t$.} \label{fig:phase-diagram}
\end{figure}

When frustration is introduced into a system the insulating state
may not be N\'eel ordered and the spin correlations may not be
strongest at $(\pi,\pi)$. Therefore, a natural question to ask is
what kinds of superconducting states do we expect to find when the
spin correlations are different from the commensurate $(\pi,\pi)$
correlations? In this Letter we study an RVB theory of the
Hubbard-Heisenberg model on the half filled ATL to investigate this
question. This is partially motivated by the fact that this model
may describe whole families of organic superconductors
\cite{springer-review}. We find that as we vary the frustration in
our model the peak in the spin fluctuations in the insulating state
moves continuously from $(\pi,\pi)$, characteristic of the square
lattice, via $(2\pi/3,2\pi/3)$, characteristic of the triangular
lattice, to $(\pi/2,\pi/2)$ characteristic of quasi-one-dimensional
(q1d) behaviour. This changes the symmetry of the superconducting
state (see Fig. \ref{fig:phase-diagram}),  from `$d_{x^2-y^2}$' for
weak frustration to `$d+id$' at the maximum frustration to
`$d_{xy}+s$' in the q1d regime. We argue that these effects are
realised in  organic superconductors such as \Brn, \CNn, \dmitn, and
\bI \cite{springer-review}.

\begin{figure}
\centering \epsfig{file=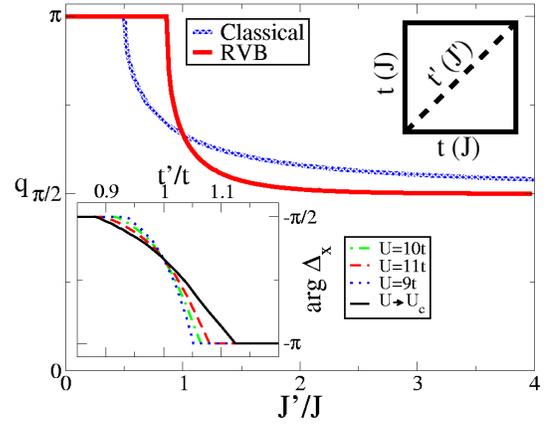, width=7cm} \caption{(Color
online.) The wavevector, $(q,q)$, where the spin correlations, which
mediate superconductivity, are strongest.  RVB theory shows that
quantum effects increase the region where commensurate correlations
are found relative to the classical theory. The lower inset shows
the angle $\theta$ which determines the symmetry of the
superconducting order parameter. 
$\theta=-\pi/2$ implies $A_2$ (`$d_{x^2-y^2}$') superconductivity;
$\theta=-\pi$ implies $A_1$ (`$s+d_{xy}$') superconductivity; and
$-\pi/2<\theta<-\pi$ implies an `$A_1+iA_2$' (`$d+id$') state. Note,
in particular, that $\theta=-2\pi/3$ for $t'=t$, independent of $U$.
The upper inset is a sketch of the ATL indicating the relevant
hopping integrals (exchange parameters) to nearest neighbours (solid
lines) and across one diagonal (dashed line).
}\label{fig:ordering}
\end{figure}

The Hamiltonian of the Hubbard-Heisenberg model is
${\cal H}
=
-t\sum_{\{ij\}\sigma}\hat{c}_{i\sigma}^\dagger\hat{c}_{j\sigma}
-t'\sum_{\langle
ij\rangle\sigma}\hat{c}_{i\sigma}^\dagger\hat{c}_{j\sigma} 
+ J\sum_{\{ij\}}\hat{\bf S}_i\cdot\hat{\bf S}_j +J'\sum_{\langle
ij\rangle}\hat{\bf S}_i\cdot\hat{\bf S}_j +
U\sum_i\hat{n}_{i\uparrow}\hat{n}_{i\downarrow}
 -\mu\sum_{i\sigma}\hat{c}_{i\sigma}^\dagger\hat{c}_{i\sigma} $
where $\hat{c}_{i\sigma}^{(\dagger)}$ annihilates (creates) an
electron on site $i$ with spin $\sigma$, $\hat{\bf S}_i$ is the
Heisenberg spin operator, and $\{ij\}$ and $\langle ij\rangle$
indicate sums over nearest and next nearest neighbours across one
diagonal respectively \cite{springer-review,Jprime} (Fig.
\ref{fig:ordering}). We study this model at exactly half-filling as
this is appropriate for the $\beta$, $\beta'$, $\kappa$ and
$\lambda$ phase organic superconductors \cite{springer-review}.
However, studies of related doped models \cite{Ogata} suggest that
the superconducting state evolves continuously upon doping.

We study this Hamiltonian via the RVB variational ansatz
\cite{AndersonRVB,ZhangPRB,RVB-organics}, $|RVB\rangle=\hat
P_G|BCS\rangle$, where $|BCS\rangle$ is the BCS wavefunction and
$\hat P_G$ is the partial Gutzwiller projector, which we treat in
the Gutzwiller approximation.  The problem reduces to solving the
BCS and Gutzwiller variational problems simultaneously. This
requires two mean-fields: $\chi_{\bf k}=\sum_{{\bf k}'}V_{{\bf
k}-{\bf k}'}\langle\hat{c}_{{\bf k}'\uparrow}^\dagger\hat{c}_{{\bf
k}'\uparrow}\rangle$ and $\Delta_{\bf k}=\sum_{{\bf k}'}V_{{\bf
k}-{\bf k}'}\langle\hat{c}_{{\bf k}'\uparrow}\hat{c}_{-{\bf
k}'\downarrow}\rangle$, where $\hat{c}_{{\bf k}\sigma}$ is the
Fourier transform of $\hat{c}_{{i}\sigma}$, and $d$ is the fraction
of doubly occupied sites. The pairing interaction,
$
V_{\bf k}=-6(1-2d)^2[J(\cos k_x+\cos k_y) + J'\cos(k_x+k_y)],$
arises from superexchange between nearest neighbours along the
square (first term) an along one diagonal (second term), see Fig.
\ref{fig:ordering}. This potential directly links the symmetry of
the superconductivity with the magnetic degrees of freedom. We
assume singlet superconductivity. It then follows from
the functional form of $V_{\bf k}$ and basic trigonometry 
that the  mean fields may be written as $\Delta_{\bf k}=\Delta_x\cos
k_x + \Delta_y\cos k_y + \Delta_d\cos(k_x+k_y)$ and $\chi_{\bf k}=
\chi_x\cos k_x + \chi_y\cos k_y + \chi_d\cos(k_x+k_y)-\tilde\mu$,
the renormalised chemical potential $\tilde\mu$ ensures
half-filling. The symmetry of the ATL is represented by the
group $C_{2h}$ 
\cite{springer-review}. A basic theorem of quantum mechanics is that
the eigenstates must transform like an irreducible representation of
group which represents the symmetry of the Hamiltonian. It follows
from this requirement that $|\Delta_x|=|\Delta_y|$,
$\chi_x^2=\chi_y^2$, and
$\theta\equiv\arg{\Delta_x}=-\arg{\Delta_y}$ \cite{confirm}.
Thus 
$\Delta_{\bf k} = |\Delta_x|\cos\theta(\cos k_x+\cos
k_y)+|\Delta_d|\cos(k_x+k_y)\notag\\+i|\Delta_x|\sin\theta(\cos
k_x-\cos k_y)$. 

For simplicity we only consider Mott insulating states that are spin
liquids, i.e., do not possess long-range magnetic order.
  We are aware that for some parameters, e.g., large $U/t$ and
small $t'/t$, that states with magnetic order may have slightly
lower energy. However, the d-wave spin liquid states considered here
are quite competitive in energy 
\cite{LeeRMP,nacoo}. Further, other work shows that the instability
of such ordered states to superconductivity, as $U/t$ decreases,
occurs for similar parameters as for spin liquid states.
 Hence, we suggest this simplifying assumption will not change our
main results relating the superconducting symmetry to the spatial
anisotropy of the spin correlations in the parent Mott insulator.

\begin{figure}
\begin{center}
\epsfig{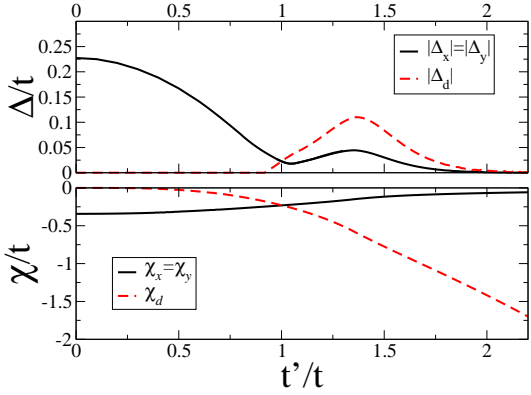}
\end{center}
\caption{(Color online.) Variation of the RVB mean-fields as the
frustration is varied for $U=10t$. $\chi_x,\chi_y,$ and $\chi_d$ all
vary smoothly, but two phase transitions occur between
superconducting states. The first occurs when $\Delta_d$ becomes
finite. But the second involves the angle $\theta$ (inset to Fig.
\ref{fig:ordering}). $\Delta_x,\Delta_y,$ and $\Delta_d$ become
small at large $t'/t$ due because the bandwidth $W\propto t'$ in
this regime and so we are moving away from the Mott transition as
$t'/t$ increases for $t'\gg t$ (c.f., Fig.
\ref{fig:phase-diagram}).} \label{fig:xi-delta}
\end{figure}

\begin{figure*}
\begin{center}
\epsfig{file=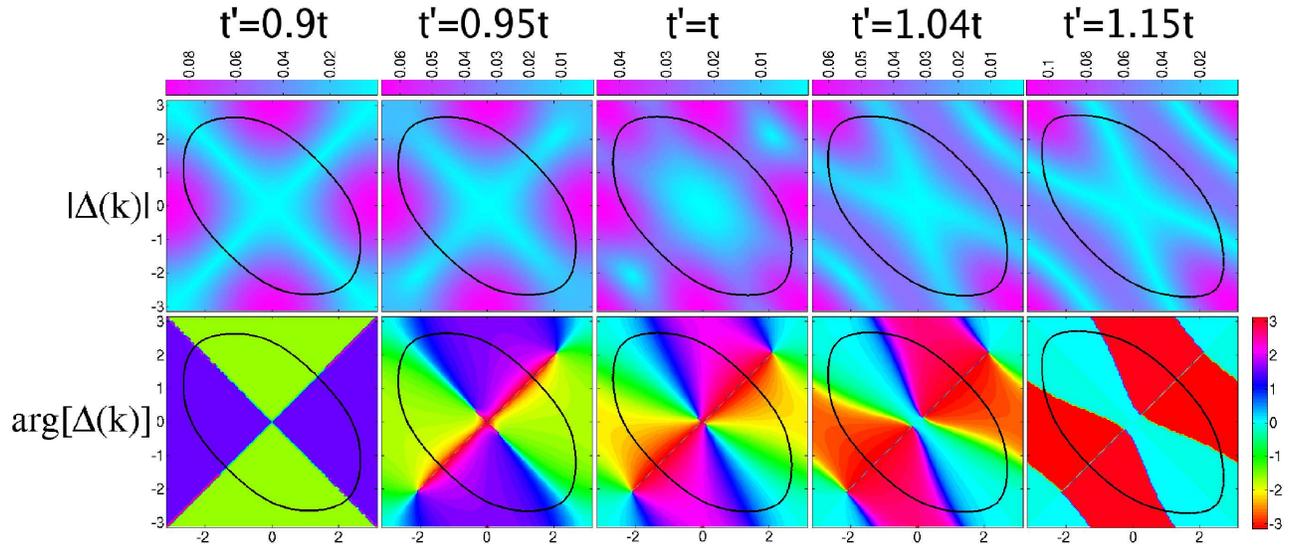,width=17cm,angle=0}
\end{center}
\caption{(Color online.) The $k$-dependence of superconducting order
parameter, $\Delta({\bf k})$ as the frustration ($t'/t$) is varied.
Each plot covers the first Brillouin zone and the solid lines denote
the non-interacting Fermi surfaces. For $t'/t<0.92$ we have a
`$d_{x^2-y^2}$' superconductor where $\Delta_{\bf k}$ transforms
as the $A_2$ representation of $C_{2v}$ 
and has symmetry required nodes along the lines $k_x=\pm k_y$. This
superconducting state is driven by the strong spin correlations at
wavevector $(\pi,\pi)$. For $t'/t>1.06$ we find an `$s+d_{xy}$'
order parameter which transforms like the $A_1$ representation of
$C_{2v}$.  This superconducting state is favoured by the strong spin
correlations near $(\pi/2,\pi/2)$. We find that this state has
nodes, although they are not required by symmetry and therefore
their location is dependent on $t'/t$ and $U/t$. For $t'=t$ the
lattice maps onto the hexagonal lattice and the ground state is a
`$d+id$' state which transforms as the $E_2$ representation of
$C_{6v}$. Intermediate states such as those found at $t'/t=0.95$ and
1.04 still show the effects of the strongly frustrated triangular
spin correlations and form `$A_1+iA_2$' states. 
The `$A_1+iA_2$' states found for $0.92<t'/t<1.06$
all break time reversal symmetry and are fully gapped. The results
shown are for $U=10t$.} \label{fig:everything}
\end{figure*}

Within the Gutzwiller approximation, $d=0$ in the insulating phase
and the model is equivalent to the Heisenberg model. Therefore,
results in the insulating phase do not explicitly depend on $U$.
However, in the insulating state the Hubbard model over-represents
the Heisenberg model. This leads to an $SU(2)$ degeneracy of the
insulating phase of the RVB theory \cite{LeeRMP,Ogata}. Physically
this means that in the Mott insulator the mean fields are not
physically distinct and the physical order parameters are
$D=\sqrt{\Delta_x^2+\chi_x^2}=\sqrt{\Delta_y^2+\chi_y^2}$ and
$D'=\sqrt{\Delta_d^2+\chi_d^2}$. It is straightforward to show that
the spin correlations are peaked at the wavevector $(q,q)$ where
$q=\arccos\left({D^2}/{2D'^2}\right).$ We solve the variational
problem numerically on an $1000\times 1000$ $k$-space mesh. Fig.
\ref{fig:ordering} compares the wavevector found in this way from
the RVB theory with the classical result,
$q=\arccos\left({J}/{2J'}\right)$ \cite{classical}. For $J'/J<0.87$
($t'/t<0.93$) we find that the spin correlations are commensurate
and peaked at $(\pi,\pi)$, consistent with a tendency towards N\'eel
ordering. We also find commensurate spin correlations [peaked at
$(\pi/2,\pi/2)$] for large $J'/J$. This is the classical ordering
wavevector for uncoupled chains. It is difficult to determine
exactly when the correlations becomes commensurate, as there is a
smooth crossover (see Fig. \ref{fig:ordering}). However, it is clear
that $q\sim\pi/2$ for $J'/J\gtrsim1.7$ ($t'/t\gtrsim1.3$). This
shows that quantum effects enhance the stability of the region with
commensurate spin correlations compared to the classical result.
This effect is also found by other theoretical methods
\cite{largeN}. In the region $0.87<J'/J\lesssim1.7$
($0.93<t'/t\lesssim1.3$) the insulating state is characterised by
incommensurate spin correlations (except at the high symmetry point
$t'=t$, see below).
 In the metallic
phase the two mean fields are physically distinct. 
$\chi_{\bf k}$ varies smoothly as the frustration, $t'/t$ is varied
(Fig.  \ref{fig:xi-delta}), but, three distinct superconducting
phases are observed as is indicated by the behaviour of $\Delta_{\bf
k}$ (Figs. \ref{fig:ordering}, \ref{fig:xi-delta}, and
\ref{fig:everything}).

For small $t'/t$ we find that $\Delta_d=0$ and $\theta=-\pi/2$, thus
$\Delta_{\bf k}=\Delta_x(\cos k_x-\cos k_y)$. This is the
prototypical form for a `$d_{x^2-y^2}$' superconductor. Formally,
$\Delta_{\bf k}$ transforms according to the $A_2$
representation of
$C_{2v}$
. This is consistent with the fact
that for small $t'/t$ the spin correlations in the insulating state
are peaked at $(\pi,\pi)$ (c.f., Fig. \ref{fig:ordering}).

For large $t'/t$ we find that $\Delta_x,\Delta_y,\Delta_d\ne0$ and
$\theta=-\pi/2$. Thus the order parameter 
takes the form $\Delta_{\bf k} = |\Delta_x|(\cos k_x+\cos
k_y)+|\Delta_d|(\cos k_x\cos k_y-\sin k_x\sin k_y)$. The first three
terms are usually referred to as `extended $s$ ($xs$)' order
parameters as they transform according to the trivial
representation, but may have accidental nodes. However, the fourth
term would be referred to as a `$d_{xy}$' state on the square
lattice. On the square lattice these `$xs$' and `$d_{xy}$' states
belong to different irreducible representations of $C_{4v}$.
We therefore refer to this state as the `$s+d_{xy}$' state. However,
we stress that $\Delta_{\bf k}$ transforms solely as the $A_1$
representation of $C_{2v}$; 
a direct consequence of the lower symmetry of the ATL. In this
regime the spin correlations are (nearly) commensurate at
$(\pi/2,\pi/2)$, as expected for weakly coupled chains. Thus it is
these q1d correlations that cause the superconductor to take
`$s+d_{xy}$' symmetry.

For $t'\sim t$; $\Delta_x,\Delta_y,\Delta_d\ne0$ and
$-\pi<\theta<-\pi/2$. Thus $\Delta_{\bf k}$ has a non-trivial
complex phase and breaks time reversal symmetry (TRS): this might be
detected by muon spin relaxation experiments \cite{group,Luke,SRO}.
The real part is the same as $\Delta_{\bf k}$ for large $t'/t$ and
transforms according to the $A_1$ representation. The imaginary part
takes the same form as $\Delta_{\bf k}$ for small $t'/t$ and
transforms according to the $A_2$ representation. We therefore refer
to this state either as the $A_1+iA_2$ or `$d+id$' state. In this
regime we have competition between  spin correlations characteristic
of the square lattice, which promote `$d_{x^2-y^2}$'
superconductivity, and those along the diagonal which favour a
`$s+d_{xy}$' state. The compromise between these frustrated
interactions is the $A_1+iA_2$ state with broken TRS. Fig.
\ref{fig:everything} details how $\Delta_{\bf k}$ varies with the
frustration for $t'\sim t$.

Exactly at  $t'=t$ the lattice becomes hexagonal and has $C_{6v}$
symmetry. In the insulating phase we find commensurate spin
fluctuations peaked at $(2\pi/3,2\pi/3)$. In the superconducting
state `$xs$' terms transform like the $A_1$ representation of
$C_{6v}$. However, the `$d_{x^2-y^2}$' and `$d_{xy}$' terms
transform as the $E_2$ representation. $E_2$ is a two-dimensional
representation spanned by the `$d_{x^2-y^2}$' and `$d_{xy}$' terms.
Thus, `$d+id$' states that transform as the $E_2$ representation are
expected on symmetry grounds on the hexagonal lattice for
appropriate values of the Ginzburg-Landau coefficients
\cite{Sigrist&Ueda,group}. This has already led to the prediction of
broken TRS in \CN and \dmit on phenomenological grounds
\cite{group}. For $t'\sim t\ne t'$ the lattice is slightly distorted
away from $C_{6v}$ symmetry and this leads to either the
`$d_{x^2-y^2}$' (for $t'\lesssim t$) or `$s+d_{xy}$' (for $t'\gtrsim
t$) component giving a greater contribution to $\Delta_{\bf k}$.
This is clearly seen in Figs. \ref{fig:ordering} and
\ref{fig:everything}. Experiments on \CCC \cite{CCC} and other
calculations \cite{largeN} suggest that RVB underestimates the size
of the region where $q\simeq2\pi/3$. As these frustrated spin
fluctuations drive $A_1+iA_2$ superconductivity this suggests that
RVB theory may underestimate the stability of this phase and the
size of the region of the phase diagram (Fig.
\ref{fig:phase-diagram}) where $A_1+iA_2$ superconductivity occurs.

We note that our phase diagram (Fig. \ref{fig:phase-diagram})
differs in the region around $t' \sim t$, from that recently
proposed by others \cite{strong-organics}. These differences arise
because those works did not
 consider the possibility of
insulating states with incommensurate spin correlations or `$d+id$'
superconducting states, and so found a spin liquid state (with
commensurate spin correlations) for $t' \sim t$ and large $U/t$.

There is significant evidence that RVB physics is enhanced on
frustrated lattices
\cite{nacoo,springer-review} and
that it is relevant to layered organic superconductors
\cite{RVB-organics,strong-organics,springer-review}.
 Further,
the RVB theory
predicts a pseudogap in the metallic
state  above the superconducting critical temperature.
Below about 50~K such a pseudogap is suggested by NMR relaxation
rate and Knight shift data \cite{NMRreview,springer-review}.
Additionally the insulating state of \CN is a spin liquid
\cite{Shimizu,springer-review}. Thus these materials provide an
testing ground for the ideas presented here. The band structures
suggest that these materials span the parameter range where the
different superconducting order parameters occur
\cite{springer-review}. For example, $t'<t$ in \Cl and \Br which
suggests that they have `$d_{x^2-y^2}$' ($A_2$) order parameters,
$t'\sim t$ in \CN and \dmit and we propose that they have `$d+id$'
($A_1+iA_2$) order parameters, and $t'>t$ in  \bI which suggests
that it has an `$s+d_{xy}$' ($A_1$) order parameter
\cite{springer-review}.  These results are consistent with our
current knowledge of the superconducting states of these materials,
but much controversy remains over the experimental situation
\cite{springer-review,group}. It has also been argued that the
superconducting state of the doped triangular lattice compound \NaH
is an RVB state with `$d+id$' pairing \cite{NCO,Ogata}.
Ferromagnetic fluctuations are strong in doped triangular lattice
systems \cite{merino} and so the possibility of triplet
superconductivity needs to be considered carefully in both doped and
half-filled systems \cite{Leggett}. 

We have studied the RVB theory of the Hubbard-Heisenberg model on
the ATL. Varying the frustration $t'/t$ changes the spatial
anisotropy of the spin correlations, which vary from being peaked on
$(\pi,\pi)$ for small $t'/t$, to incommensurate fluctuations for
$t'\sim t$ [except at $t'=t$ where the $(3\pi/2,3\pi/2)$
fluctuations are commensurate], to being commensurate at
$(\pi/2,\pi/2)$ for large $t'/t$. This drives changes in the
symmetry of the superconducting state. We propose that, as
`$d_{x^2-y^2}$' results from proximity to a N\'eel ordered state, so
`$d+id$' superconductivity arises from proximity to a spiral state
and `$s+id_{xy}$' superconductivity is driven by q1d spin
fluctuations. The generality of the connection between $(\pi,\pi)$
spin correlations and `$d_{x^2-y^2}$'
\cite{AndersonRVB,LeeRMP,Pines-Scalapino,Norman}  suggests that our
results are valid beyond the Hamiltonian studied and the
approximations used in this Letter. This clearly begs the question:
which superconducting states are driven by proximity to other
magnetic orderings? 

We thank J. Barjaktarevic, R. Coldea, J. Fj\ae restad, J. Merino, R.
Singh, S. Sorella, and E. Yusuf for stimulating conversations. This
work was funded by the ARC.


\begin{thebibliography}{999}
\bibitem{Jprime}
In principle $J=4t^2/U$ and $J'=4t'^2/U$. Our mean-field treatment
does not fully capture the renormalisation of the bare parameters.
Thus, we fix $J=t/3$ and $J'=t'^2/3t$, which correspond roughly to
the values of $J$ and $J'$ extracted from experiments on the
insulating phase of the layered organics
\cite{springer-review,Shimizu} and ensures that $J'/J=(t'/t)^2$.
\bibitem{confirm}
These identies were previously observed numerically \cite{Ogata}.
\end{thebibliography}
\end{document}